\documentclass[12pt,english]{article}
\usepackage[T1]{fontenc}
\usepackage[latin9]{inputenc}
\usepackage{amstext}

\makeatletter

\usepackage{amstext}

\makeatletter

\AtBeginDocument{\DeclareFontEncoding{T2A}{}{}}

\makeatletter

\AtBeginDocument{\DeclareFontEncoding{T2A}{}{}}

\usepackage{amsfonts}

\makeatletter

\AtBeginDocument{\DeclareFontEncoding{T2A}{}{}}

\usepackage{amsfonts}

\newtheorem{theorem}{Theorem}

\newtheorem{lemma}{Lemma}

\makeatother
\date{}

\makeatother

\makeatother

\usepackage{babel}
\begin{document}

\title{The laws of Newton and Coulomb as information transmission by virtual
particles }

\author{Malyshev V. A. \thanks{Malyshev Vadim (malyshev@mech.math.msu.su), Lomonosov Moscow State University, Faculty of Mechanics and Mathematics, Russia, 119991, Moscow, Leninskie Gory, 1}}
\maketitle
\begin{abstract}
In elementary particle physics the philosophy of virtual particles
is widely used. We use this philosophy to obtain the famous inverse
square law of classical physics. We define a formal model without
fields or forces, but with virtual particle - information transmitter.
This formal model admits very simple (school level) interpretation
with two classical particles and one virtual. Then we prove (in a
mathematically rigorous way) that the trajectories in our model converge
to standard Newtonian trajectories of classical physics. 
\end{abstract}

\section{Introduction}

The main result is formulated in the title of the paper. Although
it has connections with various fields of mathematics and mathematical
physics, I could not find a unique best framework for it Therefore,
this needs serious comments, that we do in this introduction.

\paragraph{Dynamical systems theory}

We consider here the formal system of recurrent equations (\ref{Delta_t}-\ref{Delta_w}),
which defines strongly non-linear dynamical system in three dimensional
space. Similar iterations of rational functions were studied by many
authors, and are normally sufficiently difficult \cite{Beardon}.
However, our problems are different, and we do not use the results
of this big science. The goal of this paper is to get closed results
for some scaling of parameters.

\paragraph{Classical particle physics }

Gravitational and electric forces, that describe so different physical
phenomena, surprisingly have the same form - inverse square law, differing
only by constant factors. These laws are related to harmonic functions
and the Poisson equation. Already long ago there exist other - geometric
- approaches to the Kepler laws, see for example \cite{Arnold,Hall}.
But more important, it appeared possible to deduce the Newton and
Coulomb laws from more general (and more complicated) physical theories
- general relativity theory of Einstein and Maxwell's electrodynamics,
correspondingly, where the fields play basic role. Here we show connection
of these laws with quite different models.

\paragraph{Numerical methods}

Computational methods in physics are quite developped, and it might
seem that the formal system (\ref{Delta_t}-\ref{Delta_w}) is just
an example of such computational schemes (for the Newton equations).
However, this is not quite true, for two reasons:

1. one needs the number of steps of order $c$ to reach times of order
1, and the main parameter $c$ can be too large;

2. but the main difference is that the time steps $\Delta_{n}t$ form
one recurrent system with coordinates and velocities.

\paragraph{About information transmission in classical and quantum physics }

In non relativistic classical physics one has the action at a distance
principle. For example, any particle with non-zero mass or non-zero
charge influences on all space around itself. Otherwise speaking,
information about this particle there is at the same time moment at
any space point. In relativistic quantum physics Lienard-Wiechert
potentials produce the same effect but with time delay.

In non-relativistic quantum mechanics (Schrodinger equation) the particles
are quantum, but the fields are classical. That is why only fields
act on the particles (action at a distance) but not vice-versa. About
possibility of interpretation of quantum mechanics as information
transmission see in \cite{Kantor}.

From another side, philosophy of modern quantum field theory removes
the difference between fields and particles, At the same time the
Hamiltonians have terms of only two types - either defining the free
movement of particles or (ultralocal interaction at one space point)
instantaneous ``chemical'' reaction - transformation of one group
of particles into another.

However, in elementary particle physics another principle is widely
used - any two particle interaction uses a virtual particle - information
transporter (called photon, graviton, ...). However, such conception
has not yet gained the status of rigorous mathematics. In particular,
the contradiction with the energy conservation law is explained with
``energy-time'' uncertainty principle, also beyond any athematical
formulations. And the natural question arises whether it is possible
to extend the idea of virtual particles - information transporters
- on classical physics.

We demonstrate such possibility, at least in one-dimensional case.
Namely, the particle trajectories in classical (non-relativistic)
physics, which, as it is common to believe, are governed by the fundamental
laws of Newton and Coulomb, can be obtained without introducing fields
and forces, but only using particles - real and virtual. The interpretation
of this model is extremely simple (on the school level), but the formulas
look sufficiently difficult. That is why the proof demands direct
analysis of these formulas as well. Moreover, surprisingly, the basis
of this model is a simple relation between energy (or force) and time,
see e.g. formula (\ref{f_n}).

The interpretations of this system (\ref{Delta_t}-\ref{Delta_w})
cannot be now verified with any experiment. But the corollaries are
the well-known physical laws. Concerning such situation the well-known
physicist G. 't Hooft wrote in \cite{hooft}: ``Physicists investigating
space, time and matter at the Planck scale will probably have to work
with much less guidance from experimental input than has never happened
before in the history of Physics. This may imply that we should insist
on much higher demands of logical and mathematical rigor than before
...''. That is why the mathematical rigor obviously was the main
goal of this paper.

\paragraph{On the information transmission in information computer networks}

Cellular automata, computer, communication, neural etc. networks are
normally presented as a graph with the set $V$ of vertices. In any
vertex $v$ of this graph there is an automaton, characterized at
any time moment $t$ by its state - the vector $x_{v}(t)$. Time is
normally discrete and simultaneous transformation of all vectors $x_{v}(t)$
is defined by some functions 
\[
x_{v}(t+1)=f_{v}(x_{w}(t),w\in V)
\]
providing the automata the information (prescription) what to do at
the next moment. An important example are neural networks where the
state of any neuron (vertex of the graph, or a particle on our language)
can be either $1$ or 0.

This principle of discreteness and simultaneity is often too rough
approximation. Even in the habitual life (post, telegraph, telephone
etc.) the information is transferred directionally through definite
canal, and the time of information arrival is very important. Not
only because of the queues but also because of the necessary synchronization,
see for example \cite{Alur-Dill}.

On the other hand, many concrete physical phenomena (for example in
kinetics) are modeled as a system of cellular automata - deterministic
and stochastic, see for example \cite{Schiff}. Here the physical
laws are, as though, approximated by automata systems, either giving
a new class of numerical methods, or a class of beautiful mathematical
analogies. All these aspects are reviewed in the fundamental monograph
\cite{Ilachinski}, see also \cite{Schiff,Kantor}.

At the same time, in theoretical physics one can see frequent attempts
to construct fundamental physical theories with discrete space-time,
or even without any space-time. And thus a network graph seems to
be the unique alternative to space, and time does not flow abstractly
and independently of life, but depends on the state of the network.
Although our model admits various interpretations with common one-dimensional
space and standard time, but in in fact the time for the main particle
emerges as a corollary of the state of all system.

It is important to note that, during the deduction of inverse square,
law the dimension 3 has never been used. At the same time to deduce
it from Maxwell equations (or Poisson equation) the dimension 3 is
absolutely necessary.

As a hypothesis we also note that our model can probably be generalized
for any dimension and any number of particles.

\section{Model and the main result}

\paragraph{Formal definition of the model}

The model, that we denote by $M_{0}(c,\gamma)$, has two parameters
$c>0,\gamma>0,$ and three sequences of numbers $t_{n},w_{n},y_{n},n=0,1,2,...$,
defined by the initial conditions 
\begin{equation}
t_{0}=0,y_{0}\neq0,w_{0}\label{initial_cond}
\end{equation}
 and non-linear recurrent relations (we denote $\alpha=\gamma c^{-2}$)
\begin{equation}
t_{n+1}-t_{n}=\Delta_{n}t=\frac{2y_{n}+\alpha}{c}(1-\frac{w_{n}}{c})^{-1}\label{Delta_t}
\end{equation}

\begin{equation}
y_{n+1}-y_{n}=\Delta_{n}y=\alpha+w_{n}\Delta_{n}t=\alpha+w_{n}\frac{2y_{n}+\alpha}{c}(1-\frac{w_{n}}{c})^{-1}\label{Delta_y}
\end{equation}
\begin{equation}
w_{n+1}-w_{n}=\Delta_{n}w=\frac{2\alpha}{\Delta_{n}t}=\frac{c\alpha(1-\frac{w_{n}}{c})}{y_{n}+\frac{\alpha}{2}}\label{Delta_w}
\end{equation}

\paragraph{Interpretations}

We give a simple interpretation of the formal model as a particle
movement along the real line. But firstly we define the simplified
model.

At time moments $t\in[0,\infty)$ on the real line there is the fixed
particle 1 with coordinate $z_{1}(t)\equiv0$ and the particle 2 with
coordinate $z_{2}(t)=y(t)>0$. There is also the virtual particle
$0$, interaction transporter, with coordinate $z_{0}(t)$. It is
assumed that for all $t$ 
\[
z_{1}(t)\equiv0\leq z_{0}(t)\leq z_{2}(t)=y(t)
\]
 Let us assume also that initially at time $t_{0}=0$ 
\[
z_{2}(0)=z_{0}(0)=y_{0}>0
\]
The particle $0$ runs back and forward in-between the particles $1$
and $2$ with constant sufficiently large velocity $c$, reflecting
from any of them, moreover the coordinate of the particle 2 changes
unevenly. During all other time the particle 2 stands still. Denote
\begin{equation}
0=t_{0}<t_{1}<...<t_{n}<...,\label{time_moments}
\end{equation}
 the subsequent moments, when the particle $0$ collides with particle
$2$, that is when $z_{0}(t)=z_{2}(t)=y(t)$. At these moments the
coordinate of particle 2 gets the increment $\alpha>0$, that is 
\[
y(t_{n})=y(t_{n}-0)+\alpha,
\]
Thus\c{ }we get the following system of recurrent equations for the
variables $t_{n},y_{n},n\geq0,$ 
\begin{equation}
\Delta_{n}y=y_{n+1}-y_{n}=\alpha,\,\,\,\,\Delta_{n}t=t_{n+1}-t_{n}=\frac{2y_{n}+\alpha}{c}\label{simplest_Delta_t}
\end{equation}
 that has evident solution 
\[
y_{n}=y_{0}+n\alpha,\,\,\,t_{n}=\frac{2y_{0}+\alpha(1+2n)}{c}
\]
But more important is that if the particle 2 on the time interval
$[t_{n},t_{n+1})$ moved under influence of the potential field $E=\frac{\gamma}{y}$,
then, starting from the point $y=y_{n}$ with zero velocity, it would
pass the distance $\alpha$ for the time $\Delta t=\Delta_{n}t$,
defined from the equation 
\[
(F(y)+O(\Delta y))\frac{(\Delta t)^{2}}{2}=\alpha
\]
From this and from (\ref{simplest_Delta_t}) we get:

1) explicit formula for the force and the energy, having newtonian
or coulombian form, 
\[
F(y)\sim\frac{2\alpha}{(\Delta t)^{2}}=\frac{2\alpha c^{2}}{(2y_{n}+\alpha)^{2}}\sim\frac{\gamma}{2y_{n}^{2}}\,\,\Longrightarrow\,\,E\sim\frac{\gamma}{2y_{n}}
\]

2) time-energy relationship 
\begin{equation}
E\Delta t\sim\frac{\gamma}{c}\label{energy_time}
\end{equation}
which could be possibly related to the time-energy quantum uncertainty
relation.

This explanation of the inverse square law works only on very small
time intervals, and to prove convergence of this model to newtonian
trajectory on large time intervals, one needs to introduce velocity
of the particle 2 as the effective velocity inherited (the traversed
distance to the time necessary for this) from the movement on the
previous time interval. Thus, the velocity gets at any time $t_{n}$
positive increment defined by formula (\ref{Delta_w}). At other time
moments one can assume the velocity $w(t)$ of particle 2 constant
and equal to $w_{n}$ on all interval $[t_{n},t_{n+1})$. The coordinate
of particle 2 increases linearly on this interval. At the moments
$t_{n}$ the functions $y(t)$ and $w(t)$ will be assumed right continuous.

Then it is evident that 
\[
\Delta_{n}t==\frac{2y_{n}+\alpha+w_{n}\Delta_{n}t}{c}
\]
 from where the formula (\ref{Delta_t}) follows. The factor 2 in
the formula (\ref{Delta_w}) for the velocity jump could produce some
ambiguity. In fact, we can consider that the increment $\alpha$ of
the coordinate on the previous step appeared not by itself, but as
a result of linear increase of the velocity increment (from $0$ to
$\frac{2\alpha}{\Delta_{n}t}$) on the previous step. But the following
interpretation is even more useful.

This interpretation differs from the previous one only by the dynamics
inside the intervals $(t_{n},t_{n+1})$. Moreover, $y(t)=z_{2}(t)$
and $w(t)$ become continuous, that is there is no jump $\alpha$
at times $t_{n}$ and no jump of the velocity, However, we define
the acceleration $a(t)$ with jumps at times $t_{n}$, piece-wise
constant, left continuous, constant and equal to $a_{n}$ on the interval
$[t_{n},t_{n+1})$.

The parameter $\alpha$ still exists, but will have another meaning.
Namely, we assume that particle 2 on the time interval $(t_{n},t_{n+1})$,
having constant acceleration $a_{n}$ (or constant force $f_{n}=a_{n}$),
has to pass additional distance $\alpha$ for the time $\Delta_{n}t=t_{n+1}-t_{n}$,
then 
\[
a_{n}\frac{(\Delta_{n}t)^{2}}{2}=\alpha\Longrightarrow a_{n}=\frac{2\alpha}{(\Delta_{n}t)^{2}}
\]
At the same time the velocity will gradually change from $w_{n}$
to 
\[
w_{n+1}=w_{n}+\Delta_{n}w,\Delta_{n}w=a_{n}\Delta_{n}t
\]
Thus, we modify the recurrent formulas (\ref{Delta_t}-\ref{Delta_w})
as follows 
\begin{equation}
\Delta_{n}t=\frac{2y_{n}+w_{n}\Delta_{n}t+a_{n}\frac{(\Delta_{n}t)^{2}}{2}}{c},\label{Delta_t_a}
\end{equation}
\begin{equation}
\Delta_{n}y=w_{n}\Delta_{n}t+a_{n}\frac{(\Delta_{n}t)^{2}}{2},\label{Delta_y_a}
\end{equation}
\begin{equation}
\Delta_{n}w=a_{n}\Delta_{n}t,\label{Delta_w_a}
\end{equation}
 
\begin{equation}
a_{n}=\frac{2\alpha}{(\Delta_{n}t)^{2}}\label{a_a}
\end{equation}
 that after excluding the acceleration gives the same formulas (\ref{Delta_t}-\ref{Delta_w}).
But now, from these formulas we get in addition the expression for
the force on the interval $(y_{n},y_{n+1})$ 
\begin{equation}
f_{n}=a_{n}=\frac{2\alpha}{(\Delta_{n}t)^{2}}=\frac{2\gamma}{(2y_{n}+\alpha)^{2}}(1-\frac{w_{n}}{c})^{2}=\frac{\gamma}{2y_{n}^{2}}+D_{1}c^{-1}\label{f_n}
\end{equation}
 
\[
D_{1}=\frac{2\gamma}{(2y_{n}+\alpha)^{2}}(-2w_{n}+\frac{w_{n}^{2}}{c})-c\gamma\alpha\frac{4y_{n}+\alpha}{2y_{n}^{2}(2y_{n}+\alpha)^{2}}
\]
which has, up to $O(\frac{1}{c})$, has inverse square form and indicates
on the relation with newtonian dynamics.

Possible modification of the model could be to introduce an additional
restriction - kinetic energy conservation in collisions of particles
2 and 0. But it seems that such restriction does not give much new.

Note that similar approach in two particle or even $N$ particle case
does not necessitate detailed space structure, but only the distances
between each pair of particles. This brings together physics and systems
of interacting automata, see (\cite{Alur-Dill,Ilachinski,Kantor,Schiff}).

In a possible future theory (with discrete space-time or even without
any space-time) $\alpha,c>0$ could be fundamental constants such
that $\alpha>0$ is sufficiently small\c{ } and $c$ is sufficiently
large. 

We will prove that our model converges to the newtonian trajectories
of the particle 2 in the scaling limit 
\begin{equation}
\alpha=\gamma c^{-2},c\to\infty\label{scaling}
\end{equation}

\paragraph{Asymptotic behavior}

The following result shows that the velocity of particle 2 does not
exceed $c$, thus the virtual particle always catches up particle
2,

\begin{lemma}

If $|w_{0}|<c$, then there exists a unique solution of the system
(\ref{Delta_t}-\ref{Delta_w}) with finite $t_{n},w_{n},y_{n}$ for
all $n>0$. Moreover, $|w_{n}|<c$ for all $n$.

\end{lemma}

The following result describes qualitatively the asymptotic behavior
of the model $M_{0}(c,\gamma)$ as $n\to\infty$.

\begin{lemma}\label{lem_asymp_infty_1} For any fixed $\gamma,y_{0}>0,w_{0}\geq0,,$
and sufficiently large $c>0$ (that is for $c>c_{0}$ for some $c_{0}=c_{0}(\gamma,w_{0},y_{0})$)
we have:

1) $w_{n}<c$ for all $n$;

2) the sequences $t_{n},y_{n},\Delta_{n}t,\Delta_{n}y$ are strictly
increasing to infinity as $n\to\infty$. The sequence $\Delta_{n}w$
strictly decreases to zero, and sequence $w_{n}$ increases to some
finite limit $w_{\infty}\leq c$.

If $\gamma,y_{0}>0,w_{0}<0$, then

3) there exists $N<\infty$ such that $w_{N}>0,y_{N}>0$, and for
$n<N$ we have: $y_{n}>0$ and decrease, $t_{n}$ increase, $w_{n}<0$
and increase.

\end{lemma}

However, convergence to classical mechanics occurs on the scales,
where $n$ is of the order $c$.

\begin{lemma}\label{lem_bound} For any $y_{0}>0,w_{0}\geq0,n\leq Ac,$
there exist constants $B_{i}=B_{i}(y_{0},w_{0},A)>0,i=1,2,$ such
that uniformly in $n\leq Ac$, we have the following upper bounds
\begin{equation}
t_{n},y_{n},w_{n}\leq B_{1}<\infty\label{B_1}
\end{equation}
\begin{equation}
\Delta_{n}t,\Delta_{n}y,\Delta_{n}w\leq B_{2}c^{-1}\label{B_2}
\end{equation}
 and the following lower bounds for any $n>Ac$ 
\begin{equation}
\Delta_{n}t\geq\frac{2y_{0}}{c}\Longrightarrow t_{n}\geq2y_{0}A\label{A_t_lower}
\end{equation}
 
\begin{equation}
y_{0}\leq y_{n},\alpha+w_{0}\frac{2y_{0}}{c}\leq\Delta_{n}y\label{A_y_lower}
\end{equation}
 
\begin{equation}
w_{0}\leq w_{n},\frac{2\alpha}{\Delta_{n}t}\leq\frac{2\alpha c}{B_{2}}=\frac{2\gamma}{B_{2}c}\leq\Delta_{n}w\label{A_w_lower}
\end{equation}
 Assume now $w_{0}<0$. Then the integer $N$, defined in lemma \ref{lem_asymp_infty_1},
has the bound $N\leq B_{4}c$, where 
\[
B_{4}=\frac{|w_{0}|}{\gamma}(y_{0}+\frac{\alpha}{2})
\]
 For $n>N$ the dynamics is similar to the case of positive initial
velocity with initial data at $n=N$.

\end{lemma}

\paragraph{Main result}

Consider also the trajectory $x(s)\in R$ of the newtonian particle
(interacting with another particle fixed at zero), defined by the
equation 
\[
\frac{d^{2}x}{dt^{2}}=\frac{\gamma}{2x^{2}}
\]
 and initial data 
\[
x(0)=y_{0}>0.v(0)=\frac{dx}{ds}(0)=w_{0}
\]
We assume $\gamma>0$ (repulsion). Then as $s\to\infty$ (monotonically
for $w_{0}\geq0$) 
\[
x(s)\to\infty,v(s)\to\sqrt{v^{2}(0)+\frac{\gamma}{x(0)}}
\]

\begin{theorem}

For given $c$ let $y(t)=y(t,c)$ be arbitrary curve such that for
any $n$ and any $t_{n}\leq t\leq t_{n+1}$ 
\[
y_{n}\leq y(t)\leq y_{n+1}
\]
Then for any interval $I=[0.A]$ there exists $B=B(A)$ such that
for sufficiently large $c>c_{0}(A)$

\[
|y(t)-x(t)|\leq Bc^{-1}
\]
uniformly in $t\leq t_{[Ac]}$. Then of course for any $t$ and as
$c\to\infty$ 
\[
y(t)\to x(t)
\]
 uniform\d{l}y on any finite interval.

Moreover for any function $w(t)$ such that $w_{n}\leq w(t)\leq w_{n+1}$
for any $n,t$,

\[
|w(t)-v(t)|\leq Bc^{-1},w(t)\to_{c\to\infty}v(t)
\]

\end{theorem}

\section{Proofs }

\paragraph{Proof of lemmas 1 and \ref{lem_asymp_infty_1}}

1) Note first that from formulas (\ref{Delta_t})-(\ref{Delta_w})
it follows that for $w_{0}\geq0$ all increments $\Delta_{n}t,\Delta_{n}y,\Delta_{n}w$
are positive while $w_{n}<c$. Moreover, also the sequences $t_{n},y_{n}$
are increasing.

Assume then $y_{n},w_{n}>0$ for given $n$ and let us show that the
following inequality cannot occur 
\[
w_{n}<c\leq w_{n+1}
\]
Putting $w_{n+1}=c+\epsilon,\epsilon\geq0$, from (\ref{Delta_w})
we have 
\[
w_{n+1}=c+\epsilon=w_{n}+\frac{c\alpha-\alpha w_{n}}{y_{n}+\frac{\alpha}{2}}
\]
 from where we have 
\[
w_{n}=\frac{(c+\epsilon)(y_{n}+\frac{\alpha}{2})-c\alpha}{(y_{n}+\frac{\alpha}{2}-\alpha)}=\frac{c(y_{n}-\frac{\alpha}{2})+\epsilon(y_{n}+\frac{\alpha}{2})}{(y_{n}-\frac{\alpha}{2})}=c+\frac{\epsilon(y_{n}+\frac{\alpha}{2})}{y_{n}-\frac{\alpha}{2}}
\]
This gives contradiction: if $\epsilon>0$, then $w_{n}>c$, and if
$\epsilon=0$, then $w_{n}=c$. By induction we get that this holds
for all $n$.

2) Also the sequence $w_{n}$ is increasing, then $\Delta_{n}t$ also
increase by formula (\ref{Delta_t}), and $\Delta_{n}y$ increase
by (\ref{Delta_y}). Then $t_{n}$ and $y_{n}$ tend to infinity as
$n\to\infty$. The sequence $\Delta_{n}w$ is decreasing by formula
(\ref{Delta_w}) and because $\Delta_{n}t$ increase, it tends to
zero via last equality in (\ref{Delta_w}). That is why $\Delta_{n}t\to\infty$.
It follows that $w_{n}$ tends to a finite limit by property 1).

3) If $w_{0}<0$, let $N$ be such that $y_{n}>0,w_{n}<0$ for any
$n<N$, and one of the following conditions holds:

a) $y_{N}>0,w_{N}>0$:

b) $y_{N}\leq0$;

c) $N=\infty$, that is $y_{n}>0,w_{n}<0$ for all $n$.

Let us prove that case b) is impossible. Note first that $\Delta_{n}t>0$,
and so $\Delta_{n}w>0$. while 
\begin{equation}
y_{n}>0,w_{n}<0\label{negative_w_n}
\end{equation}
that follows from formulas (\ref{Delta_t}) and (\ref{Delta_w}) correspondingly..
Then $w_{n}$, under these conditions, increases (decreases in absolute
value). Then let 
\[
y_{n+1}=y_{n}+\alpha+w_{n}\frac{2y_{n}+\alpha}{c}(1-\frac{w_{n}}{c})^{-1}<0
\]
 or 
\[
y_{n}(1+\frac{2w_{n}}{c}(1-\frac{w_{n}}{c})^{-1})<-\alpha-\frac{w_{n}\alpha}{c}(1-\frac{w_{n}}{c})^{-1}=-\alpha(1+\frac{w_{n}}{c}(1-\frac{w_{n}}{c})^{-1})
\]
which is impossible as the right hand side part is negative (as from
$|w_{0}|<c$ it follows that $|w_{n}|<c$), and the left hand side
(as $y_{n}>0$).

Let us prove now that case c) is impossible. There can be only three
cases:

c1) $y_{n}$ decrease, or more generally, tend to some limit, or even
more general, for some sequence $n_{k}$ the numbers $\Delta_{n_{k}}t$
are uniformly bounded from above by a positive constant. The $\Delta_{n_{k}}w$,
by (\ref{Delta_w}), are bounded from below by some positive constant.
Then the velocity becomes non-negative on some step. Such case has
already been considered.

c2) As we have shown, $w_{n}$ tend to some limit $w$. Firstly, let
$w<0$. Then $\Delta_{n}w\to0$, that implies $\Delta_{n}t\to\infty$
and, by (\ref{Delta_y}), $\Delta_{n}y$ asymptotically behaves like
$w\Delta_{n}t$, and then $y_{n}$ will become negative, that is impossible
as we have shown earlier.

c3) $w_{n}$ tend to the limit $w=0$. Then, as $0<\Delta_{n}w<|w_{n}|$,
\begin{equation}
\Delta_{n}y=\alpha+w_{n}\Delta_{n}t=\alpha+2\alpha\frac{w_{n}}{\Delta_{n}w}\leq-\alpha\label{less_minus_alpha}
\end{equation}
from where we get that $y_{n}$ will become negative, that is impossible.

Let us finish now the proof of point 3). In case $a)$ it was shown
above what occurs after the step $N$ when $w_{N}$ became positive.
Let us see now what was before this. We have $\Delta_{n}t>0$ by (\ref{Delta_t}),
it follows that $t_{n}$ increase, and $\Delta_{n}\text{w>0}$ by
(\ref{Delta_w}). At the same time $y_{n}>0$ by definition and decrease.
As by (\ref{less_minus_alpha}) $\Delta_{n}y<-\alpha$, thus $y_{n}$
decrease.

\paragraph{Proof of Lemma \ref{lem_bound}}

Firstly consider the case $w_{0}\geq0$. We shall get first an upper
bound for $w_{n}$ linear in $A$. As $\Delta_{n}w$ decrease, then
from the last equality of the formula (\ref{Delta_w}) it follows
that for $n\leq Ac$ 
\begin{equation}
w_{n}=w_{0}+\sum_{k=0}^{n-1}\Delta_{k}w\leq w_{0}+Ac\max_{k=0}^{n-1}\Delta_{k}w\leq w_{0}+Ac\Delta_{0}w\leq B_{3}=w_{0}+\frac{A\alpha c^{2}}{y_{0}}=w_{0}+\frac{A\gamma}{y_{0}}\label{A_w}
\end{equation}
 where the last inequality follows from the last equality of the formula
(\ref{Delta_w}).

Formula (\ref{Delta_t}) gives evident linear lower bound for $t_{n}$.
And moreover

\[
\Delta_{n}y=\alpha+w_{n}\Delta_{n}t\leq\alpha+B_{3}\Delta_{n}t\leq\alpha+\frac{2B_{3}}{c}y_{n}=
\]
 
\[
=\frac{2B_{3}}{c}(\frac{\alpha c}{2B_{3}}+y_{0}+\Delta_{0}y+...+\Delta_{n-1}y)\leq\frac{2B_{3}}{c}(1+\frac{2B_{3}}{c})(y'_{0}+\Delta_{0}y+...+\Delta_{n-2}y))\leq
\]
 
\begin{equation}
\leq\frac{2B_{3}}{c}y'_{0}(1+\frac{2B_{3}}{c})^{n}\leq\frac{2B_{3}y'_{0}}{c}(1+\frac{2B_{3}}{c})^{Ac}\leq\frac{2B_{3}y'_{0}}{c}e^{\frac{A}{2B}}\label{A_Delta_y}
\end{equation}
 where we denoted $y'_{0}=\frac{\alpha c}{2B_{3}}+y_{0}$. Then 
\begin{equation}
y_{n}=y_{0}+\Delta_{0}y+...+\Delta_{n-1}y\leq2B_{3}y_{0}e^{\frac{A}{2B}}\frac{n_{3}}{c}\label{A_y}
\end{equation}
 
\begin{equation}
\Delta_{n}t\leq\frac{2y_{n}}{c}\Longrightarrow t_{n}\leq2y_{n}A\label{A_t_upper}
\end{equation}

Consider now the case of negative initial velocity $w_{0}<0$. Let
$N$ be the first $n$, when $w_{n}$ becomes positive. Then from
the lower bound for $\Delta_{n}w$ 
\[
\Delta_{n}w=\frac{c\alpha(1-\frac{w_{n}}{c})}{y_{n}+\frac{\alpha}{2}}\geq\frac{c\alpha}{y_{n}+\frac{\alpha}{2}}\geq\frac{c\alpha}{y_{0}+\frac{\alpha}{2}}=\frac{1}{c}\frac{\gamma}{y_{0}+\frac{\alpha}{2}}
\]
 we get $N\leq B_{4}c$. where 
\[
B_{4}=\frac{|w_{0}|}{\gamma}(y_{0}+\frac{\alpha}{2})
\]
 Note also that from (\ref{less_minus_alpha}) we know that $\Delta_{n}y\leq-\alpha,$
and moreover it is easy to see that 
\[
\Delta_{n}t\leq\frac{2y_{0}+\alpha}{c},t_{N}\leq B_{4}(2y_{0}+\alpha)
\]
\[
0\leq w_{N}\leq\Delta_{N-1}w
\]

\paragraph{Bundle of Hamiltonian systems}

Further on it we will introduce some functions denoted by $D_{i}=D_{i}(y_{0},w_{0},c,n),i=1,2,...,10$,
which depend on $y_{0},w_{0},n,c$. We will se that there exist constant
$c_{0}=c_{0}(y_{0},w_{0})>0$ such that these functions are uniformly
bounded for given $y_{0},w_{0}$ and all $c,n$ such that $n<Ac,c>c_{0}$.

The introduced dynamics is not hamiltonian, as the force $f_{n}$
depends on the initial conditions $y_{0},w_{0}$ (through $y_{n},w_{n}$).
However, we will define the bundle $H=H_{y_{0},w_{o}}$ of hamiltonian
systems, which depend on $y_{0},w_{o}$, and also of $c,\gamma$,
as of parameters. Then we shall prove that any $H_{y_{0},w_{o}}$
approximates the newtonian dynamics with the same initial data. However
for other initial data $H_{y_{0},w_{o}}$ can exhibit quite different
behavior.

First of all, define the force $f(y)$ (depending of the same initial
data) on all time interval, putting it equal to $f_{n}$ on the interval
$[y_{n},y_{n+1})$.

This force defines, for any $y\in[y_{n},y_{n+1})$, the potential
$V(y)$, velocity $w(y)$ and kinetic energy $W(y)$, as follows 
\begin{equation}
V(y)=-\int_{y_{0}}^{y}f(x)dx=-\sum_{k=0}^{n}f_{k}\Delta_{k}y-f_{n}(y-y_{n})\label{V_y}
\end{equation}
 
\[
w(y)=w_{n}+f_{n}(t-t_{n}),W(y)=\frac{w^{2}(y)}{2}
\]

\begin{lemma} 
\begin{equation}
V(y)=-\frac{\gamma}{2y_{0}}+\frac{\gamma}{2y}+D_{2}c^{-1}\label{V_y_explicit}
\end{equation}

\end{lemma}

Proof. Let us rewrite (\ref{V_y}), using (\ref{a_a}) as (\ref{Delta_t}),
as 
\[
V(y)=-\sum_{k=0}^{n}a_{k}\Delta_{k}y+D_{3}c^{=1}=-\sum_{k=0}^{n}\frac{2\alpha}{(\Delta_{k}t)^{2}}\Delta_{k}y+D_{3}c^{=1}=-\sum_{k=0}^{n}(\frac{\gamma}{2y_{n}^{2}}+D_{1}c^{-1})\Delta_{k}y+D_{3}c^{=1}=
\]
 
\begin{equation}
=-\frac{\gamma}{2}\sum_{k=0}^{n}(\frac{1}{y_{k}^{2}})\Delta_{k}y+D_{5}c^{=1}=-\frac{\gamma}{2}\sum_{k=0}^{n}(\frac{1}{y_{k}}-\frac{1}{y_{k+1}}+D_{6}c^{-2})+D_{5}c^{=1}=\label{V_y_proof_1}
\end{equation}
 
\[
=-\frac{\gamma}{2y_{0}}+\frac{\gamma}{2y}+D_{2}c^{-1}
\]
 as 
\begin{equation}
\frac{1}{y_{k}}-\frac{1}{y_{k+1}}=\frac{\Delta_{k}y}{y_{k}y_{k+1}}=\frac{\Delta_{k}y}{y_{k}(y_{k}+\Delta_{k}y)}=\frac{\Delta_{k}y}{y_{k}^{2}}+D_{6}c^{-2}\label{V_y_proof_2}
\end{equation}
 Here 
\[
|D_{3}c^{=1}|\leq a_{n}\Delta_{n}t,\,\,\,\,\,|D_{5}c^{=1}|\leq|D_{1}c^{=1}|y_{n}+|D_{3}c^{=1}|
\]
 
\[
|D_{6}c^{-2}|\leq,\,\,\,\,\,|D_{2}c^{-1}|\leq\frac{\gamma}{2}|D_{6}|c^{-2}y_{n}+|D_{5}|c^{=1}
\]

\paragraph{Proof of the theorem}

Let $s,v,x$ be the time, velocity and the coordinate of the newtonian
particle satisfying the equation 
\begin{equation}
\frac{d^{2}x}{ds^{2}}=\frac{\gamma}{2x^{2}}\label{newton}
\end{equation}
with initial data $s_{0}=0,v(0),x(0)$. Denote kinetic and potential
energy of the newtonian particle as 
\begin{equation}
T=\frac{v^{2}}{2},U=\frac{\gamma}{2x}-\frac{\gamma}{2y_{0}}\label{T_U_x}
\end{equation}
correspondingly (here $-\frac{\gamma}{2y_{0}}$ is just a convenient
constant).

We will compare dynamics of two hamiltonian systems - newtonian one
and $H_{y_{0},w_{0}}$ with the same initial conditions 
\[
y(0)=x(0)=y_{0},w(0)=v(0)=w_{0}
\]
where $t,w,y$ are the time\c{ }velocity and coordinate of the particle
in the system $H_{y_{0},w_{0}}$. We will compare times $s_{n}$ and
$t_{n}$, also velocities $v_{n}$ and $w_{n}$ at the points $y_{n}$.
Then by (\ref{V_y_explicit}) and (\ref{T_U_x}) 
\begin{equation}
W(y_{n})-W(y_{0})=V(y_{0})-V(y_{n})=U(y_{0})-U(y_{n})+D_{11}c^{-1}=-T(y_{0})+T(y_{n})+D_{1}c^{-1}\label{W_T}
\end{equation}
 and hence 
\begin{equation}
w(y_{n})=v(y_{n})+D_{8}c^{-1}\label{w_v}
\end{equation}
 From the energy conservation law we have at the point $x=x(s)$ (at
some, for a while unknown, time $s$) 
\[
v^{2}(s)+(\frac{\gamma}{2x}-\frac{\gamma}{2y_{0}})=v^{2}(0)\Longrightarrow\frac{dx}{dt}=v(x(s))=\sqrt{v^{2}(0)+\frac{\gamma}{x(0)}-\frac{\gamma}{x}}
\]
and similarly at the point $y_{n}=y(t_{n})$ at time $t_{n}$ 
\[
w(t_{n})=\sqrt{w_{0}^{2}+\frac{\gamma}{y_{0}}-\frac{\gamma}{y_{n}}+D_{9}c^{=1}}
\]
 or 
\[
s=\int_{x_{0}}^{x(s)}\frac{dx}{\sqrt{v^{2}(0)+\frac{\gamma}{x(0)}-\frac{\gamma}{x}}},t=\int_{y_{0}}^{y(t)}\frac{dy}{\sqrt{w_{0}^{2}+\frac{\gamma}{y_{0}}-\frac{\gamma}{y}+D_{9}c^{-1}}}
\]
But as $v(0)=w_{0},x(0)=y_{0}$, and we consider the equation $x(s)=y(t)$,
we get 
\begin{equation}
s=t+D_{10}c^{=1}\label{s_t}
\end{equation}
and hence 
\begin{equation}
y(t)=x(s)=x(t)+D_{11}c^{=1}\label{y_t_x_s}
\end{equation}
The bounds for $D_{i}$, which are indicated in their definition,
are simple exercises in analysis, and we omit them.


\begin{thebibliography}{1}
\bibitem{hooft}Gerard $^{'}$t Hooft. Can there be Physics without
Experiments? Challenges and Pitfalls.\H{ }Revised version of Oscar
Klein lecture. Stockholm, June 1\c{ }1999), SPIN-2000/24, Int. Journal
of Modern Physics A 16 (2001) 2895-2908.

\bibitem{Arnold}Arnold V. I. Huygens and Barrow, Newton and Hooke.
1990. Birkhauser.

\bibitem{Hall}Hall R. W., Josic K. Planetary Motion and the Duality
of Force Laws. SIAM Review, 2000, v. 42, No. 1, pp. 115-124.

\bibitem{Schiff}Schiff J.L. Cellular Automata. A Discrete View of
the World. 2008. Wiley-Interscience.

\bibitem{Ilachinski}Ilachinski A. Cellular Automata. A Discrete Universe.
2001. World Scientific.

\bibitem{Kantor}Kantor F. W. Information Mechanics. 1977. Wiley.
New York.

\bibitem{Alur-Dill}Alur Rajeev, Dill David. A theory of timed automata.
Theoretical Computer Science, 1994, 126, 183-235.

\bibitem{Beardon}Beardon A. F. Iteration of Rational Functions. Complex
analytic dynamical systems. Springer-Verlag. 1990.\end{thebibliography}
\end{document}